%% file: conference_101719.tex
\documentclass[journal]{IEEEtran}
\IEEEoverridecommandlockouts
\usepackage{cite}
\usepackage{amsmath,amssymb,amsfonts}
\usepackage{algorithmic}
\usepackage{graphicx}
\usepackage{textcomp}
\usepackage{comment}
\usepackage{xcolor}

\usepackage{url}
\usepackage{hyperref}

\def\BibTeX{{\rm B\kern-.05em{\sc i\kern-.025em b}\kern-.08em
    T\kern-.1667em\lower.7ex\hbox{E}\kern-.125emX}}
\begin{document}
\title{

Explainable Speech Emotion Recognition:\ Weighted Attribute Fairness to Model Demographic Contributions to Social Bias
}

\author{Tomisin Ogunnubi, Yupei Li, Bj\"orn Schuller}

\maketitle

\begin{abstract}

Speech Emotion Recognition (SER) systems have growing applications in sensitive domains such as mental health and education, where biased predictions can cause harm. Traditional fairness metrics, such as Equalised Odds and Demographic Parity, often overlook the joint dependency between demographic attributes and model predictions. We propose a fairness modelling approach for SER that explicitly captures allocative bias by learning the joint relationship between demographic attributes and model error. We validate our fairness metric on synthetic data, then apply it to evaluate HuBERT and WavLM models fine-tuned on the CREMA-D dataset. Our results indicate that the proposed fairness model captures more mutual information between protected attributes and biases and quantifies the absolute contribution of individual attributes to bias in SSL-based SER models. Additionally, our analysis reveals indications of gender bias in both HuBERT and WavLM.
\end{abstract}

\begin{IEEEkeywords}
social bias, fairness, emotion recognition, self-supervised learning, speech analysis
\end{IEEEkeywords}
\input{introduction.tex}
\input{related_work}

\input{methodology.tex}

\input{setup.tex}
\input{results.tex}

\input{conclusion}

\bibliographystyle{vancouver}
\bibliography{refs}


\vspace{12pt}

\end{document}

%% file: introduction.tex
\section{Introduction}
Speech Emotion Recognition (SER) is the task of recognising 
emotions from speech signals. SER is vital for affective computing \cite{sun2024towards} and has applications in sectors such as mental health, education, and human-computer interaction\cite{RefWorks:elsayed2022speech, RefWorks:liu2019prototype, RefWorks:tanko2022shoelace}. 
Though existing SER models show promising results \cite{li2025large, li2025gatedxlstm}, similar to other machine learning (ML) models, SER systems are susceptible to social biases, which can systematically disadvantage specific social groups \cite{RefWorks:slaughter2023pre-trained, RefWorks:koenecke2020racial, RefWorks:caliskan2017semantics}.
This is especially a concern in scenarios where  SER is used in decision-making pipelines such as health resource allocation, hiring, or customer service interactions \cite{RefWorks:obermeyer2019dissecting, RefWorks:lee2019exploring, RefWorks:fabris2025fairness}. Biased predictions in these contexts can lead to ``allocative harm'' where individuals from certain demographic groups are denied access to resources or services, as categorised by Suresh et al.\ \cite{RefWorks:suresh2021framework}.


Fairness-aware SER research has largely focused on bias mitigation \cite{RefWorks:gorrostieta2019gender, RefWorks:lin2024emo-bias: } and bias evaluation is performed using group fairness metrics such as Equal Opportunity \cite{RefWorks:hardt2016equality} or Statistical Parity (equal distribution of positive predictions across groups)\cite{RefWorks:kusner2017counterfactual}.
While useful, these metrics typically assess protected attributes in isolation and fail to capture their joint relationship with model errors, limiting their ability to detect intersectional biases. More expressive metrics could improve bias detection, provide better interpretability, and offer stronger objectives for future mitigation efforts.

Therefore, we propose Weighted-Attribute Fairness (WAF), a new metric for fairness evaluation in SER models that learns model fairness in terms of protected attributes, capturing these joint and absolute fairness quantities that traditional metrics fail to reflect.

In this study, we make the following contributions. We propose Weighted-Attribute Fairness (WAF) and validate it on a synthetic dataset with structured bias, demonstrating that it achieves a high correlation (0.82) with the ground-truth mutual information, used as a proxy for fairness, thereby outperforming Equal Opportunity and Statistical Parity. We further evaluate the impact of incorporating both demographic and non-demographic features when modelling bias. In addition, we apply WAF to assess HuBERT \cite{RefWorks:hsu2021hubert:} and WavLM \cite{RefWorks:chen2022wavlm:} models fine-tuned on CREMA-D \cite{RefWorks:cao2014crema-d:}, and find indications of sex bias. Finally, we show that WAF provides accurate estimates of absolute error contributions by attribute, which is integrated into fairness-aware training objectives.

%% file: related_work.tex
\section{Related Work}

\subsection{Traditional Fairness Metrics}
Traditional group-based fairness metrics assess whether a model treats different groups within a protected attribute (e.g gender) fairly by comparing its predictions across those groups. \cite{Refworks:caton2024fairness, RefWorks:mehrabi2021survey}. Commonly used metrics include Demographic or Statistical Parity (SP) \cite{RefWorks:kusner2017counterfactual}, Equal Opportunity (EO) \cite{RefWorks:pessach2022review, RefWorks:hardt2016equality}, and False Positive Rate (FPR). 
These metrics are traditionally designed for binary classification tasks, and applying them in multi-class classification typically involves binarising each class: treating the target class as positive and all others as negative, then computing the metric per class \cite{RefWorks:czarnowska2021quantifying, RefWorks:pagano2023bias}. Despite their utility, some limitations exist \cite{RefWorks:zhao2022inherent, RefWorks:czarnowska2021quantifying}. Many are ratio-based and offer only a relative fairness score that compares groups for a single attribute, rather than quantifying absolute bias in the model. As a result, it is difficult to assess the severity of bias and set effective optimisation objectives. Additionally, these metrics assess sensitive attributes independently, ignoring their interactions \cite{RefWorks:chen2024fairness}. As a result, they fail to capture how attributes jointly influence prediction errors or which contribute most to overall unfairness. This lack of joint modelling makes it difficult to prioritise mitigation strategies or design fairness-aware training objectives without introducing conflicting goals, as improving fairness for one attribute may inadvertently degrade it for another.


\subsection{Social Bias in SER Models}

Social harms caused by ML systems are commonly categorised into two types: allocative harm (when a system unfairly distributes resources between different groups) and representational harm (when ML systems perpetuate harmful stereotypes against groups of people) \cite{RefWorks:suresh2021framework, RefWorks:blodgett2020language}. 
The State-Of-The-Art (SOTA) SER models rely on self-supervised learning (SSL) models trained on large speech corpora\cite{RefWorks:wagner2023dawn, RefWorks:wang2021fine-tuned, RefWorks:latif2021survey}, but still risk amplifying social biases embedded in their training data, thereby potentially causing representational harm and allocative harm when deployed in decision-making processes. Moreover, several studies have identified biases in SER systems. Gorrosieta et al.\  \cite{RefWorks:gorrostieta2019gender} demonstrated gender bias, while Lin et al.\ \cite{RefWorks:lin2024emo-bias:} attribute this largely to data bias in the downstream dataset. Speech Embeddings Associations Test (SpEAT) \cite{RefWorks:slaughter2023pre-trained} has been introduced to detect representational biases and reveal demographic biases in gender, and age.

%% file: methodology.tex
\section{Methodology}

\subsection{WAF Model Architecture}

To address the above gaps in existing fairness metrics, particularly their limited attention to attribute-level contributions, we propose a model, WAF, designed to quantify the downstream fairness of a SER model by learning how fairness varies as a function of demographic factors.
\begin{figure}[h]
    \centering
    \includegraphics[width=0.8\linewidth]{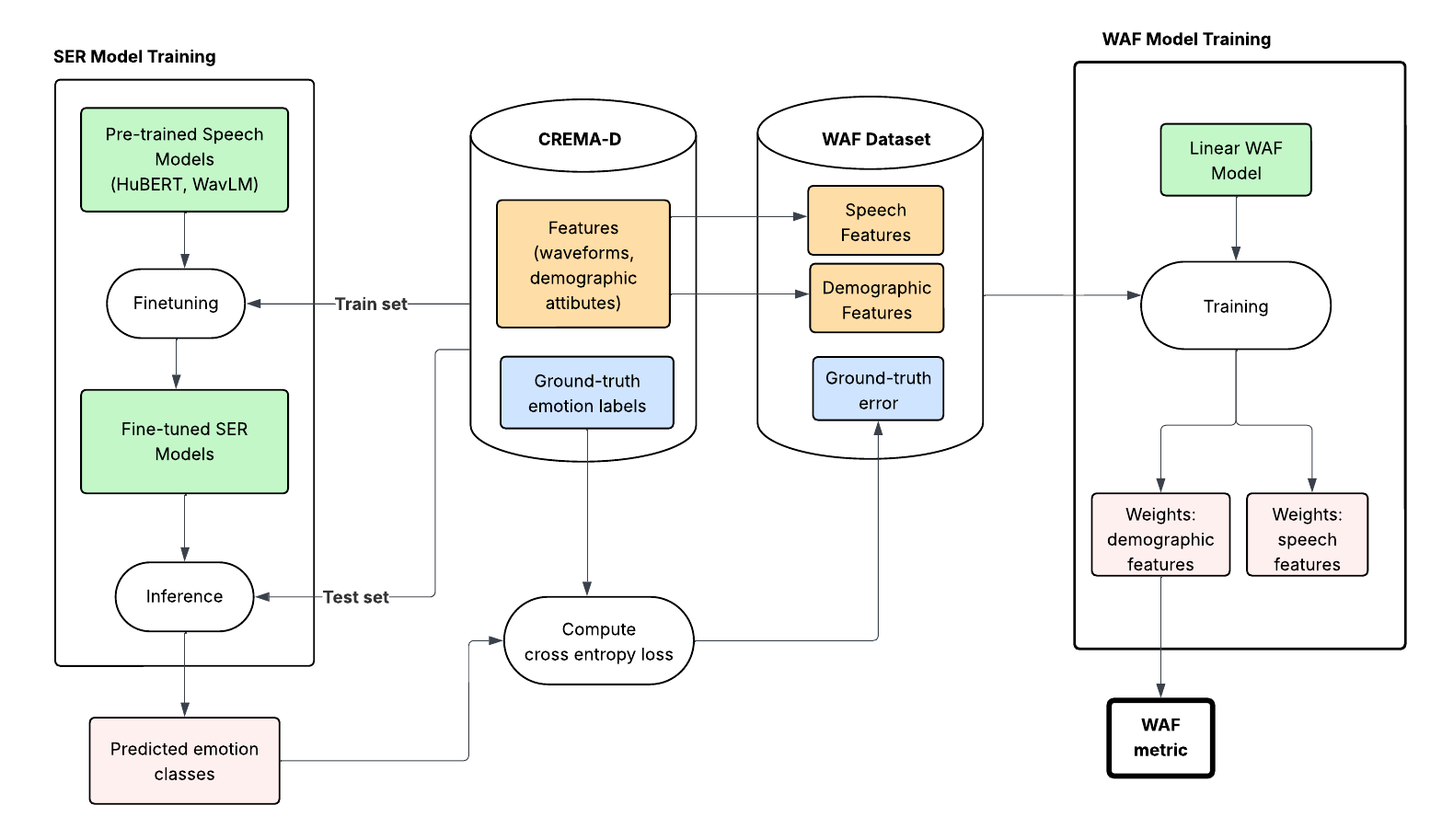}
    \caption{Pipeline for Training and Evaluating WAF Models from SER model outputs. Pre-trained SSL speech models (HuBERT and WavLM) are first fine-tuned on the CREMA-D dataset for emotion classification in SER model training phase. Their predictions are compared with ground-truth labels to compute class-specific errors. These class-level errors are then used as targets for the WAF model training, while the corresponding demographic and speech features serve as input variables. A linear model is subsequently trained to predict the class-level errors from these features. The learned coefficients associated with demographic attributes define the WAF metric, quantifying both the magnitude and direction of each attribute’s contribution to model error and fairness.}
    \label{fig:waf-architecture}
\end{figure}

In our work, we define fairness in terms of allocative error, calculated as the loss incurred by the SER model on each sample for each emotion. We design the pipeline as seen in Fig.\ \ref{fig:waf-architecture} and formulate it as follows: 
%
 %
 
For a sample, let the allocative error $\mathbf{e}_i$ be defined as, $\mathbf{e}_i = -(\mathbf{y}_i \odot \log(\hat{\mathbf{y}}_i)
    + (1 - \mathbf{y}_i) \odot \log(1 - \hat{\mathbf{y}}_i)) $,
where: $\mathbf{y}_i\in \{0,1\}^{|C|}$ is the one-hot vector of ground-truth labels for the $|C|$ emotion classes, $\mathbf{\hat{y}}_i\in [0,1]^{|C|}$ is  predicted softmax probabilities and $\odot$  denotes element-wise product.

This is the binary cross-entropy loss computed per class, where each term penalises the model for assigning low probability to the true class and high probability to the false classes. By decomposing the error per class, this approach allows fairness to be assessed individually across emotion classes.





To train the WAF model, we use both demographic and speech features from the CREMA-D dataset. Each input sample $\mathbf{x_i}$ is a concatenation of the two feature types. Demographic features are binarised into privileged (1) and unprivileged (-1) categories, allowing fairness direction to be tracked \cite{RefWorks:verma2018fairness, RefWorks:czarnowska2021quantifying}, while speech features are selected empirically.

WAF is implemented as a simple feed-forward neural network with a single hidden layer and an output layer projecting to six outputs with ReLU activation. The model predicts the allocative error per class, $\hat{\mathbf{e}}_i = \text{WAF}(\mathbf{x}_i; \theta) $, and it is trained with minimise mean squared error.

The hyperparameters used are available on
\href{https://github.com/myxp-lyp/Explainable-Speech-Emotion-Recognition}{our GitHub}. Upon training completion, we interpret the learnt weights associated with the demographic variables $\theta_d$ as indicators of their impact on fairness. These serve as the `WAF scores'. A value of 0 indicates complete fairness, i.\,e., this attribute has no impact on the error, a value $>$ 0 indicates bias against the privileged class (1), i.\,e., samples belonging to the privileged class (1) contribute more towards the error; otherwise (for value $<$ 0), the unprivileged class exhibits the bias.

\subsection{Feature Selection}
We hypothesise that non-demographic information, such as speech features, also contributes meaningfully to classification errors, alongside demographic attributes. However, it is unclear how the inclusion of such features might affect the model's understanding of demographic contributions to model performance. To explore this, we investigate whether augmenting the WAF model with speech features enhances its ability to accurately predict model error while remaining robust in isolating the influence of demographic attributes.


To investigate this, we extract a subset of features from SSL embeddings generated by fine-tuned HuBERT and WavLM models. Principal component analysis (PCA) \cite{pearson1901pca} is applied to the embeddings, and we focus on the first principal component (PC1), which captures the greatest variance. We examine the loadings of PC1 and select the top-k dimensions with the largest absolute contributions, forming a reduced feature subset. These selected dimensions are then concatenated with demographic features to serve as model input.


We evaluate the performance of WAF by computing the mean squared error (MSE) with varying $k$ to study how incorporating speech features impacts predictive ability. The case where $k = 0$ corresponds to a baseline model using demographic-only features, allowing us to measure the incremental performance gains from using speech features. 
Additionally, we use a mean regressor baseline, which predicts the global average of the target variable for all samples to establish a lower bound for performance.


\subsection{Validation with Synthetic Data}


WAF is designed as a reliable fairness metric capable of identifying disparities between groups and their magnitudes. To validate it, we create a synthetic CREMA-D dataset with controlled demographic disparities across emotions, providing a ground-truth reference. Then we evaluate how well WAF and other metrics correlate with this ground truth. A high correlation ($>0.7$) indicates effective capture of attribute-driven bias.

To simulate SER outputs, each sample is assigned softmax probabilities over six emotion classes. We introduce six biased attribute–emotion mappings (Table \ref{tab:error_contribution}) by injecting bias into samples from privileged groups. For a selected emotion, privileged samples receive a low probability (0.1–0.3) if it is the true class and a high probability (0.7–0.9) otherwise, increasing misclassification and error. Unbiased samples receive high probabilities for the true class, yielding more accurate predictions. Probabilities are then normalised to ensure a valid distribution. For example, for Anger, samples from privileged Sex and Age groups (Sex = 1, Age = 1) are assigned lower true-class probabilities, resulting in higher error, whereas for Happy, only Age contributes to the injected bias (Table \ref{tab:structured_error}).


We quantify the dependence between each protected attribute and each emotion error using Mutual Information (MI) \cite{cover2006elements}, which provides a continuous measure of linear and non-linear dependence \cite{RefWorks:brillinger2004data} and correctly isolates the attributes which affect performance for an emotion class by assigning them larger values. We report both Pearson and Spearman correlations between MI and fairness metrics (SP, EO, FPR, and WAF) across all attribute-emotion pairs. 


\begin{table}[h]
  \caption{Attribute subsets associated with injected bias in the synthetic dataset: for each emotion class, high error was injected into the privileged group of the specified attributes.}
    \centering
    \begin{tabular}{|c|c|}
     \hline
       Emotion Class & Privileged Attribute Subset \\ \hline
        Anger     & Age, Sex \\
        Disgust   & Race, Ethnicity \\
        Fear      & None \\
        Happy     & Age \\
        Neutral   & Age, Sex, Race, Ethnicity \\
        Sadness   & Age, Sex, Race \\
        \hline
    \end{tabular}
      \vspace{-0.8cm}

    \label{tab:error_contribution}
\end{table}
\begin{table}[h]
     \caption{Examples of structured error injection for A and H. For Anger, high error is introduced when both Age and Sex are in the privileged group (1), modelling an intersectional bias. For H, only Age contributes to the error. }
    \centering
    \begin{tabular}{|c|c|c|c|}
    \hline
    Age & Sex & Error (Anger) & Error (Happy)\\
    \hline
    1 & 1 & 0.9 & 0.9  \\
    1 & -1 & 0.1 & 0.9 \\
    -1 & 1 & 0.1 & 0.1 \\
    -1 & -1 & 0.1 & 0.1 \\
          \hline
    \end{tabular}

    \label{tab:structured_error}
    \vspace{-0.7cm}
\end{table}

%% file: setup.tex
\section{Experimental Setup}

To investigate SER bias across multiple protected attributes, we utilise the Crowd-sourced Emotional Multimodal Actors Dataset (CREMA-D)\cite{RefWorks:cao2014crema-d:}, which has six balanced basic categorical emotions:  Anger (A), Disgust (D), Fear (F), Happy (H), Neutral (N) and Sad (S). The data has the availability of demographic information for each speaker, including sex, age, race, and ethnicity. We split them into train and test sets with an 80:20 ratio, stratified by emotion and using a fixed random seed 42. This results in 5953 samples in the training set and 1489 samples in the test set. 

Additionally, this study focuses on SSL-based SER models. We select two widely used SSL models, HuBERT and WavLM\cite{RefWorks:hsu2021hubert:, RefWorks:chen2022wavlm:}. Both models are fine-tuned on the CREMA-D dataset using a linear classification head, consisting of a single hidden linear layer. The hyperparameters are shown in codes. Each model achieves good performance on the test set, comparable to existing benchmarks \cite{RefWorks:wu2024emo-superb:}: HuBERT: 69\% accuracy, WavLM: 62\% accuracy. These models serve as the basis for our downstream fairness evaluations.

%% file: results.tex
\section{Results}


\subsection{Fairness Evaluation of SSL-Based SER Models}
 
\begin{figure}[h]
    \centering
    \includegraphics[width=0.8\linewidth]{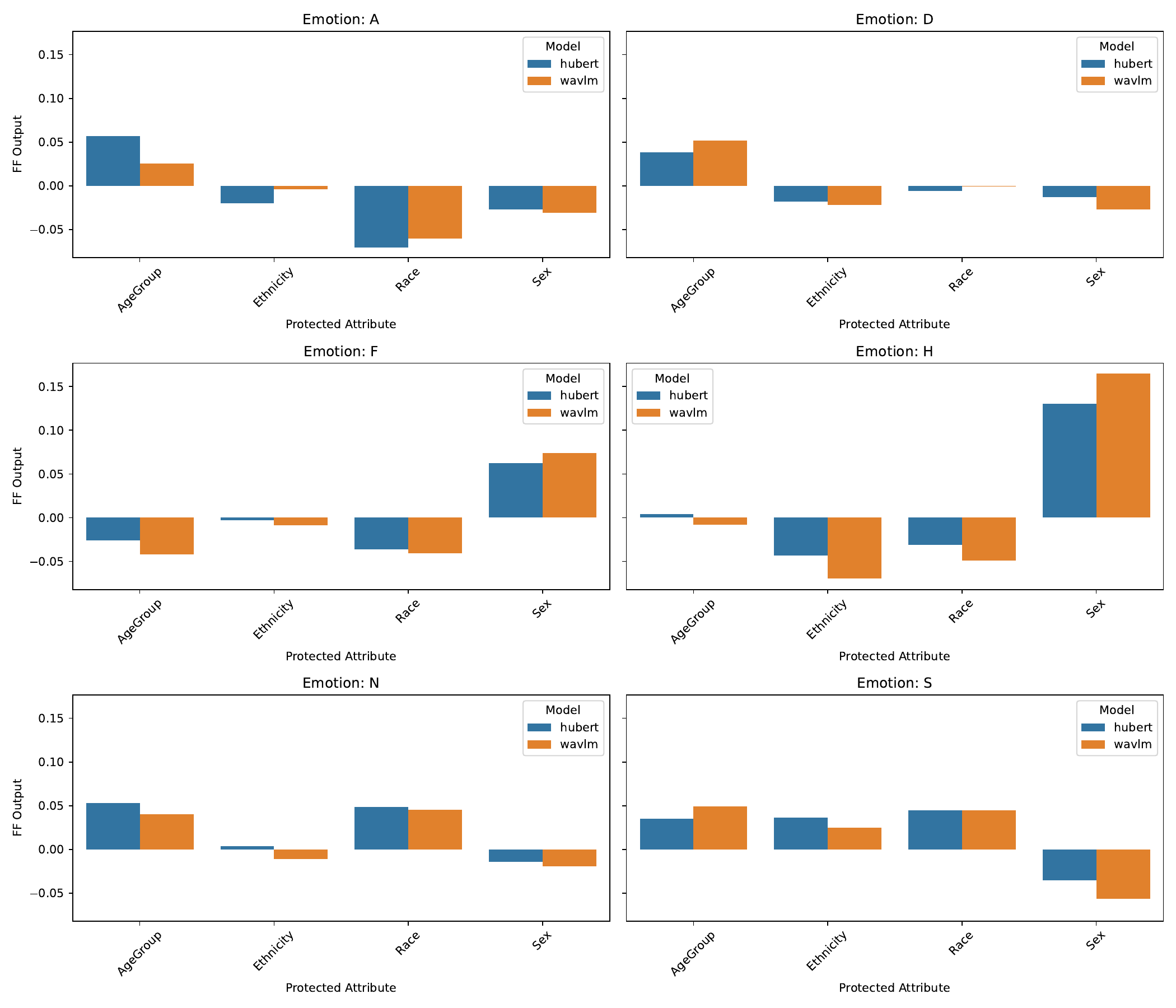}
    \caption{WAF fairness scores for the WavLM and HuBERT models, reported per emotion class and protected attribute}
    \label{fig:waf_models}
\end{figure}

Using WAF, we analyse the fairness of HuBERT and WavLM fine-tuned on CREMA-D. We define privileged groups as follows: Male (Sex), Caucasian (Race), Not Hispanic (Ethnicity), and ages 20–35 years (AgeGroup). WAF scores are learnt with respect to these groups, such that positive values indicate bias against the privileged group. Lower absolute values indicate greater fairness. Table \ref{tab:waf_score} and Fig \ref{fig:waf_models} summarise WAF scores by demographic attribute and emotion class.

\begin{table}[h]
\caption{WAF scores for HuBERT and WavLM (three decimal places). Bold values indicate the largest absolute contribution per emotion class within each model.}
\centering
\resizebox{\columnwidth}{!}{
\begin{tabular}{|c|cccc|cccc|}
\hline
& \multicolumn{4}{c|}{\textbf{HuBERT}} & \multicolumn{4}{c|}{\textbf{WavLM}} \\
\hline
& AgeGroup & Ethnicity & Race & Sex 
& AgeGroup & Ethnicity & Race & Sex \\
\hline
A & 0.041 & -0.005 & \textbf{-0.075} & -0.004 
  & 0.030 & -0.007 & \textbf{-0.060} & -0.025 \\

D & \textbf{0.047} & -0.020 & -0.004 & -0.027 
  & \textbf{0.051} & -0.027 & 0.005 & -0.023 \\

F & -0.037 & 0.002 & -0.037 & \textbf{0.055} 
  & -0.046 & 0.002 & -0.037 & \textbf{0.071} \\

H & 0.015 & -0.031 & -0.023 & \textbf{0.140} 
  & 0.001 & -0.062 & -0.041 & \textbf{0.180} \\

N & 0.071 & 0.017 & \textbf{0.080} & -0.057 
  & 0.059 & -0.005 & \textbf{0.080} & -0.058 \\

S & 0.040 & 0.041 & 0.039 & \textbf{-0.068} 
  & 0.053 & 0.036 & 0.045 & \textbf{-0.073} \\

\hline
\end{tabular}}
\label{tab:waf_score}

\end{table}

Sex bias is observed across multiple emotion classes in both models. HuBERT and WavLM mostly disadvantage the unprivileged group (female), with negative coefficients for four emotions (A, D, S, N). Given CREMA-D’s uniform emotion distribution but higher male proportion, this data imbalance may influence fairness direction. However, the largest absolute fairness scores occur against males for H and F, indicating higher misclassification for the privileged group. This may stem from representational bias in speech embeddings and warrants further study, aligning with prior findings on gender disparities in SER outputs.
Bias directionality for both the race and ethnicity attributes is largely consistent, suggesting shared underlying demographic signals in the data. 
The magnitudes remain relatively low (mostly under 0.08), indicating that race and ethnicity contribute weakly to model bias, overall. Similarly, only moderate bias is observed for AgeGroup, mostly against the privileged class (20-35 years).



WavLM demonstrates consistently higher bias magnitudes than HuBERT across nearly all emotion classes except Anger.
These results may be a wider representation of the lower accuracy of the WavLM model (62\% compared to HuBERT's 67\%) but may also suggest that WavLM may be more susceptible to encoding demographic information in its downstream predictions.



\subsection{Validating Relative and Absolute Bias Detection with WAF}
We validate WAF against traditional fairness metrics on our synthetic dataset. WAF achieves the highest Pearson correlation with MI (0.82), effectively capturing attribute contributions to error. FPR also correlates strongly (-0.79) but slightly underperforms WAF, while EO and SP show only moderate correlations, reflecting lower sensitivity to attribute-driven errors. This advantage arises from WAF’s use of full probability distributions, capturing nuances missed by argmax-based metrics. Spearman correlations are weaker for all metrics, likely due to low-impact cases.

\begin{figure}[h]
    \centering
    \includegraphics[width=0.8\linewidth]{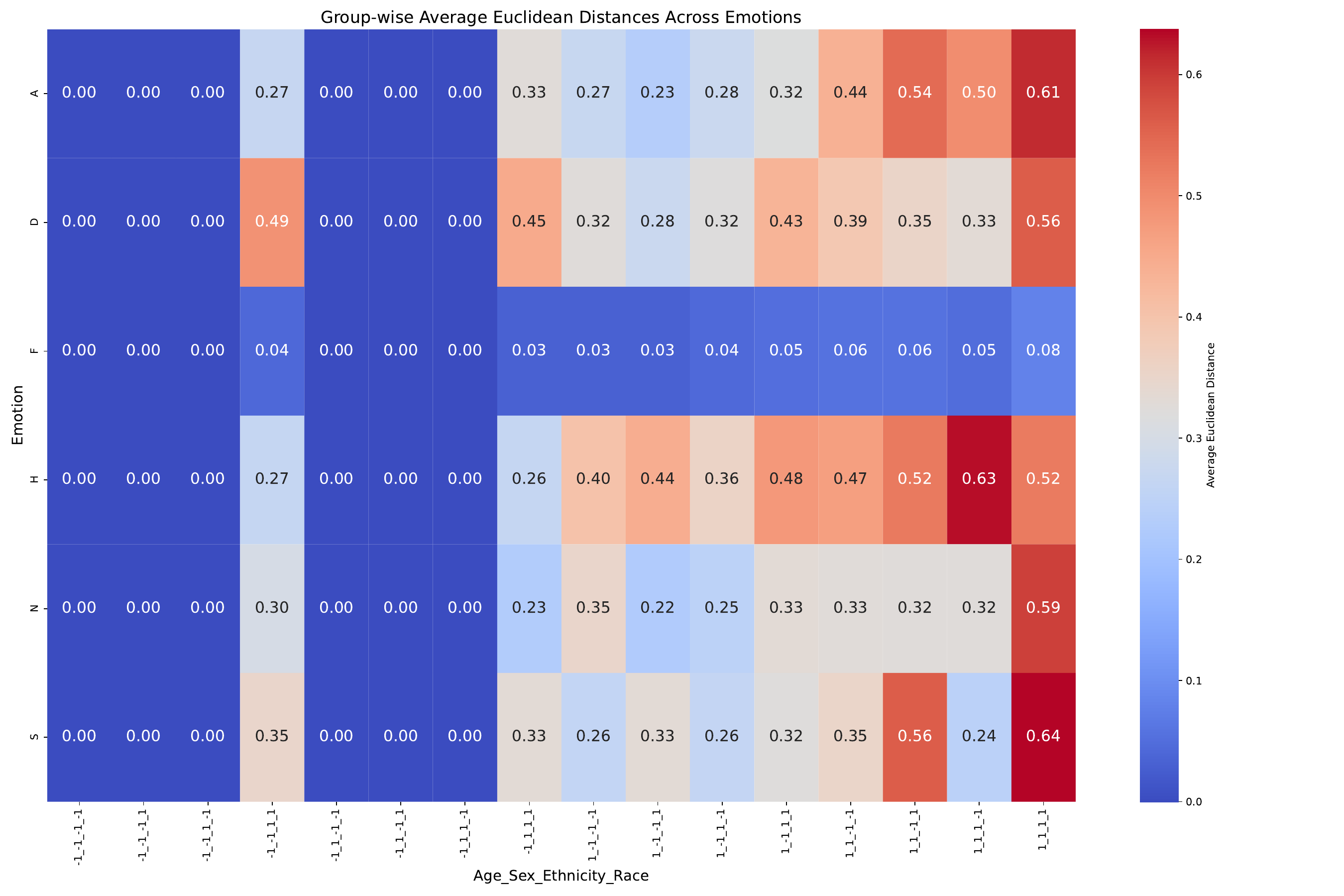}
    \caption{Average Euclidean distance between total loss estimated by WAF coefficients and the ground truth mean loss across joint combinations of demographic attributes}
    \label{fig:euc_dist}
\end{figure}

Beyond its correlation with MI, WAF also captures each attribute’s absolute contribution to total prediction error. We evaluate this by comparing ground-truth average errors for joint demographic groups (e.g., Age=1, Sex=1, Ethnicity=-1, Race=-1) with WAF’s estimates, computed as a linear combination of attribute coefficients. The Average Euclidean Distance between predicted and actual group-emotion losses is shown in Fig. \ref{fig:euc_dist}.

Distances increase with the number of privileged attributes, with group 1111 (all privileged) showing the highest deviation of 0.64, about 5\% of the dataset’s total error range (0.01–14.9), which is acceptable for interpretability. Fear shows uniformly low deviations (0.03), reflecting its lack of injected bias and low ground-truth error, which WAF accurately captures. This makes WAF suitable for bias-aware training, where minimising both total loss and demographic contributions can promote fairness without harming performance.

\begin{table}[h]
  \caption{Pearson and Spearman correlations between mutual information (MI) of attribute-emotion error, and fairness metrics. Values in \textbf{bold} have the highest magnitude. WAF shows the highest Pearson correlation. Values marked with $^*$ have $p$-values $>$ 0.05}
    \centering
    \begin{tabular}{|c|c|c|c|c|}
       \hline
    & WAF & EO & FPR & SP \\
    \hline
    Pearson & \textbf{0.82} & 0.59 & -0.79 & -0.63  \\
    Spearman & 0.6 & 0.33$^*$ & \textbf{-0.61} & -0.32$^*$\\
          \hline
    \end{tabular}
  
    \label{tab:mi_correlation}
    \vspace{-0.7cm}
\end{table}



\subsection{Impact of Feature Selection}

\begin{figure}[h]
\vspace{-0.5cm}
    \centering
    \includegraphics[width=0.8\linewidth]{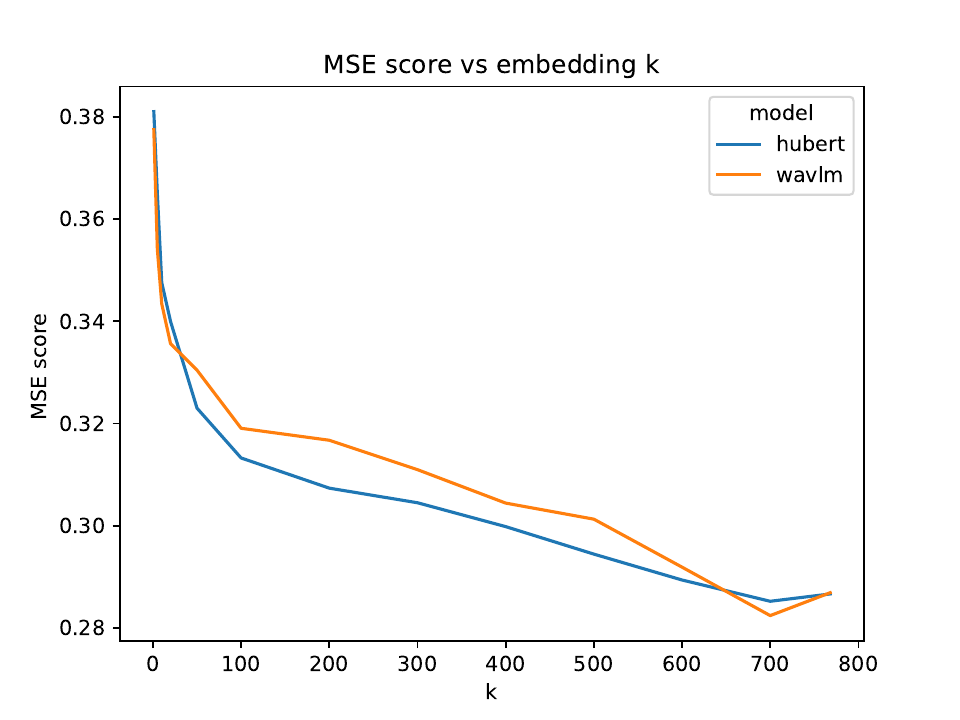}
    \caption{$MSE$ of the WAF model as the embedding dimension $k$ increases}
    \label{fig:mse_k}
        \vspace{-0.3cm}

\end{figure}

As shown in Fig.\ \ref{fig:mse_k}, the mean squared error (MSE) of the model decreases sharply with increasing $k$, particularly for small values ($<$ 50), indicating that early principal components of the speech embeddings contain highly predictive information about the target. For both the HuBERT and WavLM models, performance improves steadily, but it slows from $k=100$, suggesting diminishing returns beyond this point.

The baseline mean regressor achieves an MSE of 0.32 for HuBERT and 0.36 for WavLM. These baselines are exceeded once $k > 50$, showing that even a relatively small subset of speech features provides substantial predictive value. Accordingly, we select $k = 100$ as a practical compromise, capturing most predictive gain while keeping the model simple. This value is used in the WAF experiments. Incorporating speech-derived features captures non-demographic variability, helping to explain errors more fully, reduce reliance on demographic attributes, improve predictive accuracy, and enhance the interpretability of fairness scores for the SER models.

%% file: conclusion.tex
\section{Conclusion}
This work investigated allocative bias in SSL-based SER models and assessed whether fairness models can capture it in an interpretable way. Our proposed WAF was validated on a synthetic dataset with structured bias, and clearly identifying the most influential attributes. It could highlight the relative importance of demographic features and accurately estimate each attribute’s absolute contribution to prediction error. We also examined the role of both demographic and non-demographic features, revealing that bias is context-dependent and not purely demographic. Applying the framework to HuBERT and WavLM revealed multi-demographic biases in both models, particularly related to sex. Future work will explore bias mitigation using WAF and model representational harm in SSL-based SER systems.


